\documentclass[aps,twocolumn,prl,showpacs,preprintnumbers,superscriptaddress,amsmath,amssymb,floatfix]{revtex4}
\usepackage{graphicx}% Include figure files
\usepackage{dcolumn}% Align table columns on decimal point
\usepackage{bm}% bold math
\usepackage[english]{babel}
\usepackage{color}

\begin{document}
\preprint{APS/123-QED}

\author{C. Stingl}
\affiliation{I. Physikalisches Institut,
Georg-August-Universit\"{a}t G\"ottingen, D-37077 G\"ottingen}
\author{R.S. Perry}
\affiliation{Scottish Universities Physics Alliance, School of
Physics, University of Edinburgh, Mayfield Road, Edinburgh EH9 3JZ,
Scotland} \affiliation{International Innovation Center, Kyoto
University, Kyoto 606-8501, Japan}
\affiliation{Department of
Physics, Kyoto University, Kyoto 606-8502, Japan}
\author{Y. Maeno}
\affiliation{International Innovation Center, Kyoto University,
Kyoto 606-8501, Japan} \affiliation{Department of Physics, Kyoto
University, Kyoto 606-8502, Japan}
\author{P. Gegenwart}
\affiliation{I. Physikalisches Institut,
Georg-August-Universit\"{a}t G\"ottingen, D-37077 G\"ottingen}

\title{Symmetry-breaking lattice distortion in Sr$_3$Ru$_2$O$_7$}
%\title{{\color{red}Area-preserving lattice symmetry breaking} in Sr$_3$Ru$_2$O$_7$}
\date{%TCIMACRO{\TeXButton{Date}{\today} }
%BeginExpansion
\today%
%EndExpansion
}

\begin{abstract}
The electronic nematic phase of Sr$_3$Ru$_2$O$_7$ is investigated by
high-resolution in-plane thermal expansion measurements in magnetic
fields close to 8~T applied at various angles $\Theta$ off the
c-axis. At $\Theta<10^\circ$ we observe a very small ($10^{-7}$)
lattice distortion which breaks the four-fold in-plane symmetry,
resulting in nematic domains with interchanged $a$- and $b$-axis. At
$\Theta \gtrsim 10^\circ$ the domains are almost fully aligned and
thermal expansion indicates an area-preserving lattice distortion of
order $2\times 10^{-6}$ which is likely related to orbital ordering.
Since the system is located in the immediate vicinity to a
metamagnetic quantum critical end point, the results represent the
first observation of a structural relaxation driven by quantum
criticality.
\end{abstract}

\pacs{71.10.Hf,71.27.+a,75.30.Kz} \maketitle

Electronic nematic states in condensed matter are recently
considered quantum analogues of nematic liquid crystals which
display directional but no positional order~\cite{Kivelson}. Their
ground state is intermediate between insulating "crystals" and
isotropic metallic (Fermi) liquids. Upon forming an electronic
nematic phase, the point-group symmetry of the underlying host
crystal is spontaneously broken, resulting in anisotropic electronic
properties which could most easily be detected by charge or heat
transport measurements along different crystallographic
directions~\cite{Fradkin}. A spontaneous generation of spatial
anisotropy has %been found in transport experiments on various
%different correlated electron systems including
been found in quantum Hall systems, cuprate and iron-pnictide
high-temperature superconductors as well as in the bilayer ruthenate
Sr$_3$Ru$_2$O$_7$~\cite{FQHE,Daou,Pnictides,borzi_science_2007}. In
the case of the cuprates, the nematicity characterizes the yet
unidentified pseudogap state while for iron-pnctides it precedes the
orthorhombic lattice distortion and spin-density-wave formation.
However, as the four-fold rotational symmetry in these cases is
already broken by the lattice, the transport anisotropy arising from
electronic nematicity needs to be carefully distinguished from the
one resulting from the crystal lattice alone~\cite{Fradkin}. In this
respect, Sr$_3$Ru$_2$O$_7$ is considered as a cleaner example as its
nematic order develops out of a four-fold symmetric state and is not
accompanied by magnetic ordering or charge-density-wave formation.
In this Letter, we prove by high-resolution dilatometry that the
nematicity in Sr$_3$Ru$_2$O$_7$ is related to a spontaneous
symmetry-breaking lattice distortion, representing the first example
of a structural relaxation driven by strong electronic correlations
near a quantum phase transition.

The bilayer perovskite ruthenate Sr$_3$Ru$_2$O$_7$ has initially
attracted considerable attention due to a field-induced quantum
critical point (QCP) near 8~T ($B\parallel c$), which is associated
with itinerant-electron metamagnetism~\cite{grigera_science_2001}.
Subsequent studies on ultrahigh-quality samples revealed
thermodynamic evidence for the formation of a novel phase below 1~K,
bounded in field by two first-order metamagnetic transitions, which
intervenes in the approach of the QCP~\cite{grigera_science_2004}.
Within this phase, the electrical resistivity reaches a maximum and
becomes temperature independent, indicating a reduction of the
charge carrier mean free path. This has been ascribed to the
formation of domains resulting from a Pomeranchuk-like Fermi surface
distortion. As the magnetic field is tilted by a small angle of
$\Theta=13^\circ$ towards the ab-plane, the in-plane resistivity
develops a pronounced anisotropy within the phase, suggesting the
formation of an electronic nematic fluid~\cite{borzi_science_2007}.
The isotropic resistivity behavior at $B\parallel c$ would then
arise from the random orientation of domains which could be aligned
by an in-plane field.

The electronic structure of Sr$_3$Ru$_2$O$_7$ results from the
partially filled Ru $t_{2g}$-states hybridizing with the O $p$-
orbitals. The Fermi surface is complicated because of bilayer
splitting of multiple bands, a $7^{\circ}$ rotation of the RuO
octahedra leading to a $\sqrt{2}$ reconstruction of the Brillouin
zone and spin-orbit coupling. Six bands have been experimentally
identified~\cite{Mercure_2010}, which have different quasi
one-dimensional (1D) ($d_{yz}$ and $d_{xz}$) and quasi
two-dimensional ($d_{xy}$) orbital character. According to
tight-binding band-structure calculations, the metamagnetic
transitions and the nematic ordering result from the quasi-1D bands
which are close to van-Hove (vH) singularities~\cite{Raghu,Lee,Wu}.
The nematic state is characterized by a few percent difference in
occupation between $d_{yz}$ and $d_{xz}$ derived bands. This
(partial) "orbital ordering" breaks the four-fold rotational
symmetry of the system. However, since electrons close to a vH
singularity have a very small velocity, their influence on the
charge transport in such a multiband system is negligible. Thus, the
observed resistivity anisotropy must be related to domain wall
scattering rather than being intrinsic to the nematic
phase~\cite{Raghu}. If the domains are fully lined up, e.g. by an
in-plane magnetic field, the resistivity should become insensitive
to the nematic phase, although the nematic ordering could even be
stronger. Therefore, dilatation measurements probing the intrinsic
lattice anisotropy are a much better indicator for the nematic
phase.

The measurements were performed on a high-quality single crystal
studied previously by thermal expansion and magnetostriction along
the $c$-axis~\cite{grigera_science_2004,gegenwart_PRL_2006}. In
order to determine the length change along the $a$-axis while
applying a field along the $c$-axis, a miniaturized high-resolution
capacitive dilatometer, which can be rotated in the bore of a
superconducting magnet, has been utilized. Two parallel flat springs
exert a uniaxial pressure of about 15~bar along the measurement
direction $\parallel a$. The linear thermal expansion coefficient
$\alpha(T)$ is obtained by calculating the slope of the relative
length change in temperature intervals of 40~mK.

\begin{figure}
    \includegraphics[width=0.5\textwidth]{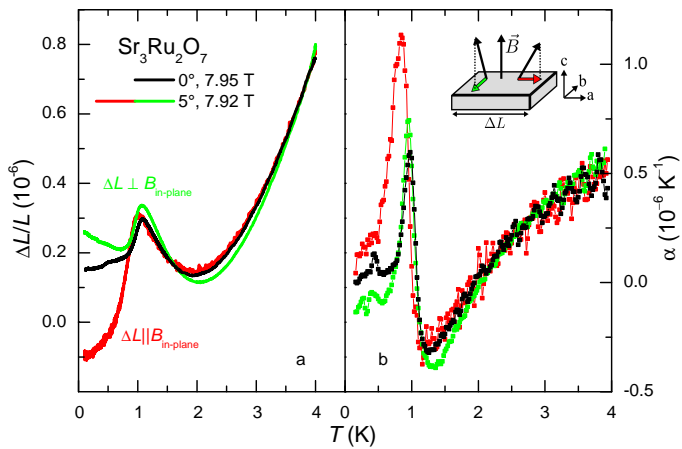}
    \caption{(color online) In-plane relative length change $\Delta L/L$ (a) and
    respective thermal expansion coefficient $\alpha=d(\Delta
    L/L)/dT$ (b) vs. temperature for Sr$_3$Ru$_2$O$_7$. The black curves represent
    data at a magnetic field of 7.95~T applied along the $c$-axis,
    whereas the green and red curves are taken at a field of 7.92~T
    tilted by $5^\circ$ from the $c$-axis with $\Delta L\perp B_{\rm in-plane}$
    and $\Delta L \parallel B_{\rm in-plane}$, as sketched in the inset.}
    \label{Fig1}
\end{figure}

At first, we focus on the thermal expansion signature of the nematic
phase for untilted magnetic fields ($B\parallel c$). As shown by the
black line in Fig.~1a, a distinct anomaly in the length change is
found around 1~K which corresponds to a sharp discontinuity in the
thermal expansion coefficient displayed in Fig. 1b, indicative of a
second-order phase transition, also found in specific heat
measurements~\cite{rost_science_2009}. Note that our previous length
measurements along the $c$-axis have only detected a very small
signature at the phase boundary~\cite{gegenwart_PRL_2006},
indicating that the nematic transition primarily affects the
$ab$-plane.

Next, we compare these data with corresponding measurements in a
tilted magnetic field. In order to investigate the $a$-axis length
change {\it parallel} to the in-plane field, the entire dilatometer
has been rotated in the bore of the superconducting magnet. In a
second run to determine the $a$-axis change {\it perpendicular} to
the in-plane field, the sample has been rotated around $a$, while
the dilatometer has been kept in its original position. In either
case, the angle $\Theta$ with respect to the $c$-axis has an error
of $\pm 1^\circ$. As shown in Fig. 1, the tilting of the field has
no significant effect at temperatures above 1.2~K. By contrast, a
pronounced anisotropy evolves at the nematic transition, which is
the central observation of this study. Despite the application of an
in-plane field, the system retains its four-fold in-plane symmetry
above the transition, which is then spontaneously broken at 1.2~K.

%The size of the phase transition anomaly
%in $\alpha(T)$ increases along the direction of the in-plane field
%and decreases along the perpendicular orientation.This points at
%the formation of domains which could be aligned by an in-plane
%field.

\begin{figure}
    \includegraphics[width=0.5\textwidth]{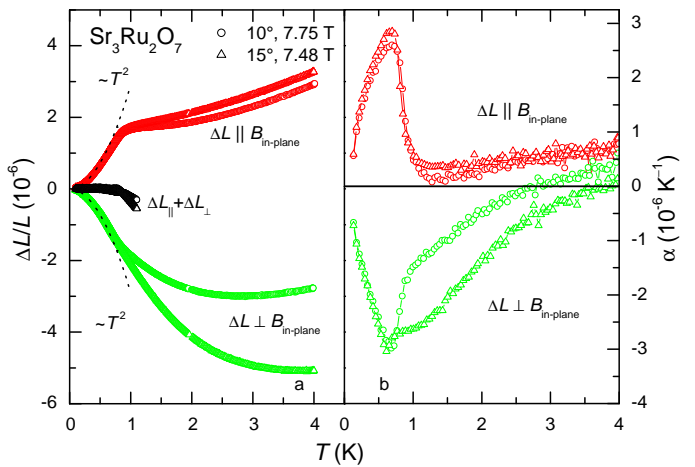}
    \caption{(color online) In-plane relative length change (a) and thermal expansion (b) at
    magnetic fields of 7.75~T (circles) and 7.48~T (triangles)
    applied $10^\circ$ and $15^\circ$ off the c-axis,
     respectively. For each angle, the length change
    has been determined parallel (red symbols) and
     perpendicular (green symbols) to the in-plane component
     of the magnetic field. The dotted lines indicate $T^2$ behavior.}
    \label{Fig2}
\end{figure}

We have also studied the lattice anisotropy in the nematic phase at
larger tilt angles, cf. Fig.~2. Compared to $\Theta\leq 5^\circ$,
thermal expansion within the phase is much enhanced and the length
changes parallel ($\Delta L_{\parallel}$) and perpendicular ($\Delta
L_{\perp}$) to the in-plane field have similar size and opposite
sign. This indicates an area preserving lattice distortion of order
$2\times 10^{-6}$. This is still smaller than the instrumental
resolution of elastic neutron scattering, which has failed to detect
a splitting of the (020) peak~\cite{borzi_science_2007}. Below 0.5~K
Fermi liquid behavior $|\alpha|=a_1T$ is found, with a highly
enhanced coefficient $a_1\approx 6\times 10^{-6}$K$^{-2}$, which is
independent of $B$ (within the nematic phase) and $\Theta$ (for
$\Theta\geq 10^\circ$).

Our experiments strongly support the domain scenario described
above. Since there is no difference in the observed distortion
between the measurements at 10$^\circ$ and 15$^\circ$, the sample
appears to be effectively monodomain at $\Theta\gtrsim 10^\circ$.
The thermal expansion discontinuities at the nematic transition
indicate that the $a$-parameter of Sr$_3$Ru$_2$O$_7$ contracts
(expands) along the direction parallel (perpendicular) to the
in-plane field upon cooling through the nematic transition. Without
an in-plane field, the uniaxial pressure parallel to the measured
$a$-axis leads to a contraction upon cooling through the nematic
transition (cf. Fig. 1), although it is not sufficient to align all
nematic domains. The contraction is enhanced by a superposed
in-plane field {\it along} the direction of the uniaxial pressure
while it is reduced for a {\it perpendicular} in-plane field.

%The important difference of the lattice anisotropy compared to the
%anisotropy of the electrical resistivity is, that the latter one
%disappears at large tilt angles as the domains are
%aligned~\cite{borzi_science_2007}, whereas the former one saturates
%representing the true nematic order parameter.

\begin{figure}
    \includegraphics[width=0.5\textwidth]{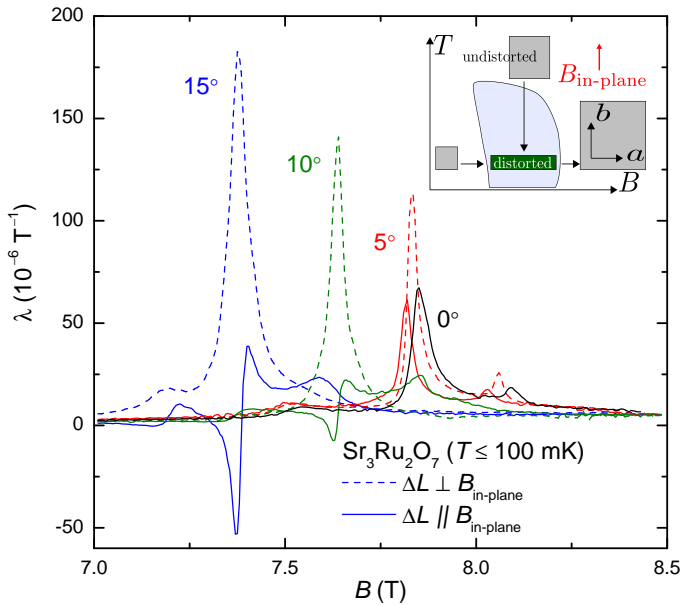}
    \caption{(color online)
    Isothermal magnetostriction at temperatures below 0.1~K
     (cf. symbols in Fig. 4) of Sr$_3$Ru$_2$O$_7$ at different
      angles $\Theta$ of the applied field to the $c$-axis.
       The solid and dotted curves represent data of the
        magnetostriction coefficient $\lambda=d(\Delta L/L)/dB$
         along and perpendicular to the in-plane field,
         respectively. The inset shows a sketch of the lattice
         distortion in the nematic phase (cf. light blue region) as
         derived from thermal expansion and magnetostriction measurements.}
    \label{Fig3}
\end{figure}

We now turn to the measurements of the isothermal magnetostriction.
At low temperatures, the nematic phase is bounded in field by two
first-order metamagnetic transitions at $B_{c1}=7.85$~T and
$B_{c2}=8.07$~T for $\Theta=0$~\cite{grigera_science_2004}. Our
previous $c$-axis dilatation measurements have detected two positive
peaks in the magnetostriction coefficient in close similarity to
respective peaks in the magnetic susceptibility, indicating a
substantial magnetoelastic coupling, leading to an expansion of the
c-axis at each increase of the sample
moment~\cite{grigera_science_2004,footnote}.

\begin{figure}
    \includegraphics[width=0.5\textwidth]{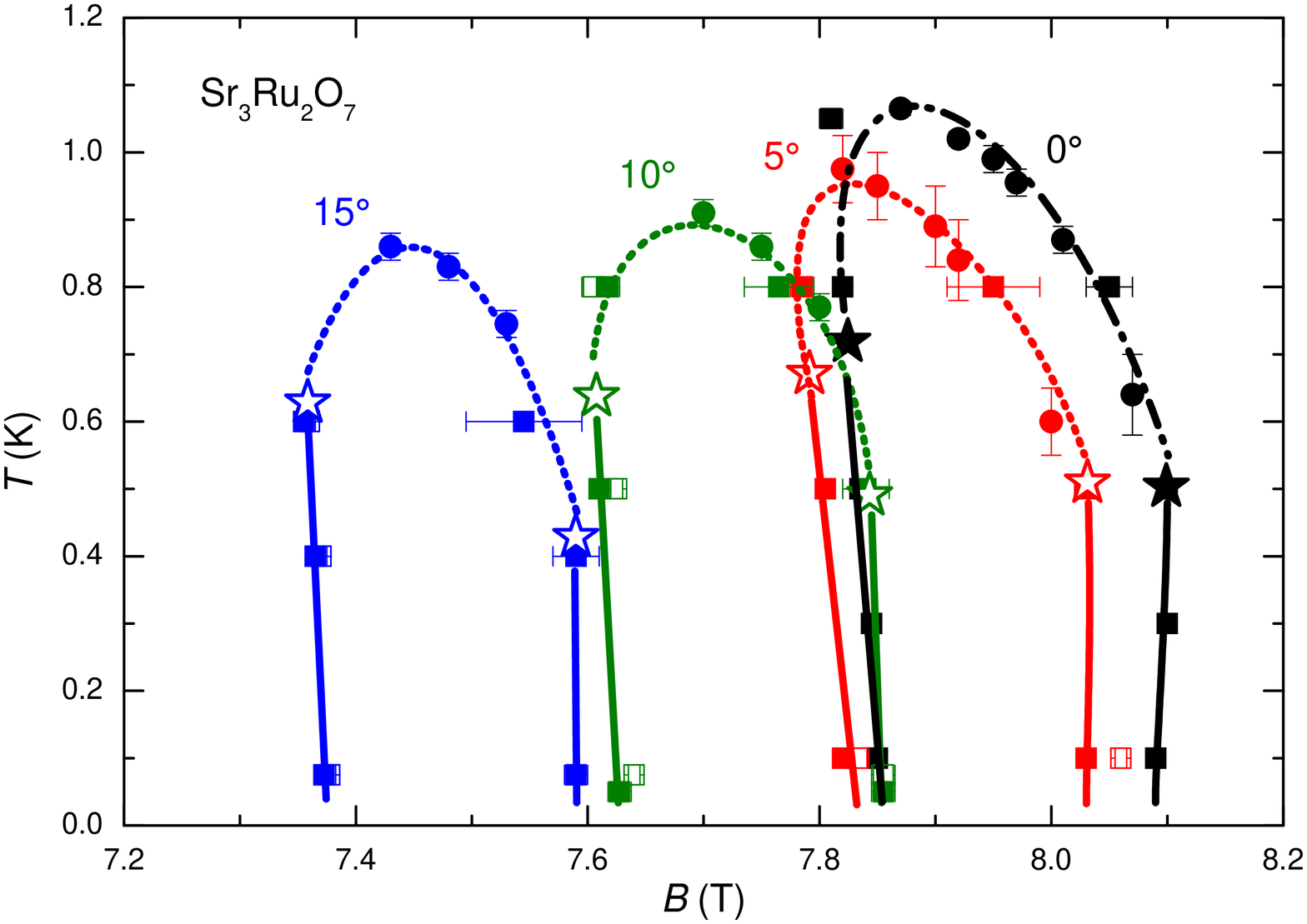}
    \caption{(color online) Phase diagram of the nematic phase of Sr$_3$Ru$_2$O$_7$
    at different tilt angles (cf. different colors). The squares and
    circles denote transitions in magnetostriction and thermal
    expansion, respectively (solid symbols: $\Delta L\parallel
    B_{\rm in-plane}$, open symbols: $\Delta L\perp B_{\rm in-plane}$). The
    black solid lines indicate first-order transitions ending at
    tricritical points (cf. filled stars)~\cite{grigera_science_2004},
    connected by a (dash-dotted) line of second-order transitions. The
    colored open stars denote the presumed position of critical points,
    connected by crossover lines (dotted).}
    \label{Fig4}
\end{figure}

The lattice distortion in the nematic phase is most clearly
reflected in low-temperature isothermal magnetostriction
measurements along the $a$-axis at angles $\Theta\gtrsim 10^\circ$,
c.f. Fig.~3. Upon entering the phase at the lower metamagnetic
transition, a pronounced expansion is found perpendicular to the
in-plane field direction, whereas parallel to the in-plane field a
contraction is observed. At $B_{c2}$ the plane then mainly expands
parallel to the in-plane field. This is compatible with an
orthorhombic distortion as sketched in the inset.

Fig.~4 displays the evolution of the phase diagram of
Sr$_3$Ru$_2$O$_7$ with the tilt angle. At $\Theta=0$, the nematic
phase is bounded in field by two lines of first-order transitions
ending at tricritical points~\cite{grigera_science_2004} that are
connected by a line of second-order phase transitions (indicated by
the dashed black line) ~\cite{rost_science_2009}. An in-plane
component of the magnetic field explicitly breaks the $C_4$
symmetry, leading to anisotropic behavior already outside the
nematic phase. The nematic transition is then no longer defined by a
spontaneous symmetry breaking. What was a nematic transition for
$\Theta=0$ should now be called meta-nematic transition in the
presence of an in-plane field~\cite{Kivelson_pr}. Specifically, the
continuous transition above the critical points should become a
crossover (cf. dotted lines), similar to the transition in an Ising
ferromagnet in a small external symmetry breaking field. The thermal
expansion signature at the "roof" of the nematic phase changes with
increasing $\Theta$, but these data alone do not allow to definitely
classify the order of the transition.
 %Indeed, the magnetostriction signatures at 100~mK
%remain sharp in tilted field, whereas the peaks in thermal expansion
%at the "roof" of the nematic phase broaden significantly at
%$\Theta\geq 10^\circ$.
%For the angular dependence of both critical fields at $T\leq 100$~mK
%we find empirically that it follows
%$B_{c,i}(\Theta)=B_{c,i}(0)\cos^2(\Theta)$.
%As indicated by the arrows, the fields at which the thermal
%expansion measurements shown in Figs. 1 and 2 were taken have been
%adjusted such that the nematic phase is always entered at the same
%relative position between $B_{c1}$ and $B_{c2}$.

The following scenario, sketched in Fig.~5, may account for our
observations: The formation of the electronic nematic phase in
Sr$_3$Ru$_2$O$_7$ leads to distorted domains with interchanged $a$
and $b$-axis, which can be aligned in tilted fields. These domains
are responsible for the peak of the electrical resistivity
\cite{grigera_science_2004}. The average domain wall separation
%, estimated from a Dingle analysis of de Haas-van Alphen
%oscillations within the nematic phase
is of the order of 500~nm~\cite{Mercure_2009}. Thermal expansion in
tilted fields indicates that the domains are almost fully aligned
for $\Theta\gtrsim 10^\circ$. However, a few percent misaligned
domains, which are hardly seen in thermal expansion, may still lead
to additional scattering in the electrical resistivity. Most
surprisingly, the resistivity peak is completely suppressed already
at $\Theta\gtrsim 5^\circ$ for the direction parallel to the
in-plane field ("easy" transport direction), whereas it remains up
to about $20^\circ$ along the orthogonal direction ("hard"
direction)~\cite{borzi_science_2007}. This indicates that the domain
wall scattering is highly anisotropic, likely because the domains
are stretched along the easy transport direction. If an in-plane
field along the in-plane $x$ direction leads to a preferential
occupation of the perpendicular $d_{yz}$ orbitals~\cite{Wu},
elongated domain walls perpendicular to the in-plane field are
energetically favorable, since they don't break the $d_{yz}$ orbital
bonds. Consequently a domain structure as sketched in Fig. 5 is
stabilized. Our measurements prove that there is an expansion of the
lattice along the easy transport direction parallel to the orbital
bonds in the major domain. The Fermi surfaces responsible for the
nematic ordering are hole like. Since electrons in hole-like bands
are similar to anti-bonding states in molecules, a reduction of the
bandwidth due to a lattice expansion is favorable, as it reduces the
kinetic energy of electrons in these states. Thus, the observed
lattice distortion is consistent with the orbital ordering
description of how four-fold rotational symmetry is broken in
Sr$_3$Ru$_2$O$_7$~\cite{Wu private}.

\begin{figure}
    \includegraphics[width=0.45\textwidth]{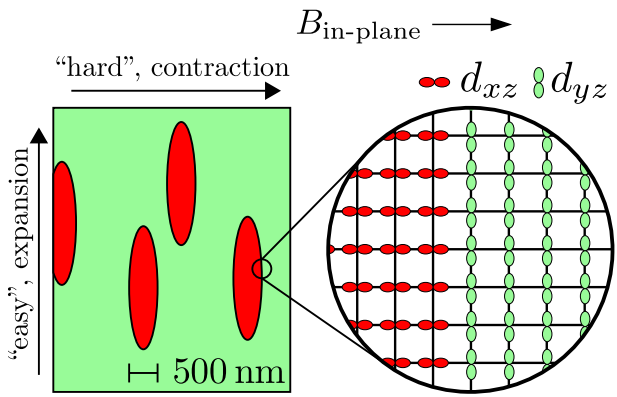}
    \caption{(color online) Illustration of the proposed domain configuration
    (left, on macroscopic scale) and orbital ordering (right, on atomic scale) for the
    nematic phase of Sr$_3$Ru$_2$O$_7$~\cite{Wu}.
    The different orbitally ordered states are indicated by the
    green and red color. The arrows indicate the directions of
    the in-plane field, easy and hard transport, as well as expansion and contraction.}
    \label{Fig5}
\end{figure}

Finally, we comment on the general implications of our observations.
Under certain conditions, fulfilled e.g. for metamagnetic or valence
instabilities, the order parameter fluctuations couple linearly to
the lattice, leading to a divergence of the
compressibility~\cite{Weickert}. Then, the lattice may become
unstable against a structural relaxation close to the QCP, as found
in our experiments on Sr$_3$Ru$_2$O$_7$. This observation should
motivate the search for other structural instabilities driven by
quantum criticality and the development of models, taking into
account the coupling between phonons and quantum critical
fluctuations.
%It has been pointed out that ultrasound experiments are well suited
%to investigate the coupling of phonons to nematic Fermi surface
%fluctuations in Sr$_3$Ru$_2$O$_7$~\cite{Adachi}.
%For the heavy-fermion system CeRu$_2$Si$_2$, which is close to but
%still off a metamagnetic QCEP~\cite{Weickert}, a 50\% softening of
%the longitudinal elastic constant has previously been found in
%ultrasound experiments~\cite{Bruls}.

In conclusion, the electronic nematic phase in Sr$_3$Ru$_2$O$_7$ is
characterized by an area-preserving symmetry breaking lattice
distortion of several $10^{-6}$. If the symmetry is not broken
explicitly by a sufficiently large in-plane field, domains with
interchanged $a$ and $b$-axis form, leading to a much smaller
overall distortion of the sample. The domain formation seems also
responsible for the observed transport
anisotropy~\cite{borzi_science_2007}. The lattice expansion
perpendicular to the in-plane field is compatible with the proposed
orbital ordering in Sr$_3$Ru$_2$O$_7$~\cite{Raghu,Lee,Wu}. In
general, our observation represents a fascinating example of a
structural relaxation driven by strong electronic correlations.
Compared to the recently discussed nematic ordering in
iron-pnictides~\cite{Pnictides}, the lattice distortion in
Sr$_3$Ru$_2$O$_7$ occurs at 100 times lower temperatures and is more
than 100 times smaller. Most remarkably it is {\it not} accompanied
by magnetic ordering and represents the first example of a lattice
instability in the immediate vicinity of a QCP.

We thank A.S. Gibbs, A.P. Mackenzie, R. K\"{u}chler and M. Garst for
collaborative work and S.A. Kivelson and C. Wu for helpful
conversations. This work is supported by the German Science
Foundation through SFB 602.

\end{document}